# Magnetic Charge State Controlled Spin-Wave Dynamics in Nanoscale Three-Dimensional Artificial Spin Ice

[1]Chandan Kumar, [2]Amrit Kumar Mondal, [1]Sreya Pal, [1]Sayan Mathur, [3]Jay R. Scott, [4]Arjen van Den Berg, [3]Adekunle O. Adeyeye, [4]Sam Ladak[*], and [1]Anjan Barman[*]

[1]Department of Condensed Matter and Materials Physics, S. N. Bose National Centre for Basic Sciences, Block JD, Sector III, Salt Lake, Kolkata 700106, India

[2]Technical Research Centre, S. N. Bose National Centre for Basic Sciences, Block JD, Sector III, Salt Lake, Kolkata 700106, India

[3]Department of Physics, Durham University, Durham DH1 3LE, United Kingdom

[4]School of Physics and Astronomy, Cardiff University, Cardiff CF24 3AA, United Kingdom

[*]E-mail: abarman@bose.res.in; ladaks@cardiff.ac.uk

Three-dimensional (3D) magnetic nanostructures offer a versatile platform for exploring complex spin textures and spin-wave (SW) dynamics, with implications in next-generation spintronic and magnonic technologies. Advances in 3D nanofabrication have allowed a wide-range of structures and phenomena to be realized. Whilst the study of simple cylindrical magnetic nanowires allows the realization of ultrafast domain walls and a spin Cherenkov effect, placing such wires of complex cross-section into 3D arrangements allows one to produce magnetic metamaterials, known as artificial spin-ice (ASI), where the overall ground state and spin dynamics are governed by magnetostatic interactions between elements. Here, using Brillouin Light Scattering (BLS) we demonstrate the direct detection of magnetic charged states in a 3D-ASI system. The measured spin-wave modes in 3D-ASI are found to be directly controlled by the local magnetic charge configuration and the direction of the applied magnetic field. Micromagnetic simulations provide insight into the spatially selective excitation of spin waves and the evolution of magnetic microstates, uncovering a direct link to the field-dependent characteristics of the spin-wave spectrum. These findings make 3D-ASI architectures a promising system to realize reconfigurable, low-power magnonic devices with engineered collective dynamics.

*Introduction*

Artificial spin ice (ASI) systems are frustrated, engineered arrays of single-domain nanomagnets with magnetostatic interactions that dictate their ordering and ground state [1]. Two-dimensional ASI (2D-ASI) systems have emerged as prime candidates for reconfigurable magnonic crystals, allowing precise control of spin wave (SW) dynamics through the manipulation of the magnetic microstates and configurational anisotropy [1-3]. The unique properties of these frustrated metamaterials have their applications in magnonics [4,5], reservoir computing [6,7], and data storage [8], along with enhancing the fundamental understanding of captivating phenomena, including emergent magnetic monopoles [9,10], vertex-based frustration [11,12], phase transitions [13], and chiral dynamics [14]. Transcending the limitations of 2D to explore the unique characteristics of these metamaterials in three-dimension (3D) is pivotal for the advancement of 3D nanomagnetism [15-17]. 3D nanomagnetism can explore different physical aspects, including geometry [16,18], topology [19], chirality [20], unconventional spin textures [21-23] and frustration [24], inter-linkage between the two or more of which may lead to exotic properties. Initially focused on investigation of static and dynamic characteristics of 3D systems such as rolled-up microtubes [25], nanowires and others[16], 3D nanomagnetism has significantly evolved with advancements in fabrication techniques. Beyond traditional processes, advanced methods like two-photon lithography (TPL)[26] and focused electron beam-induced deposition (FEBID) [27] have revolutionized the fabrication of 3D nanostructures, enabling the study of complex systems such as the diamond bond lattice [24], double-helix [20], woodpile-structure [28], nano-bridge [29] and other complex 3D structures [30]. In addition to conventional characterization techniques like magnetic force microscopy (MFM), advancements in X-ray and electron-based characterization techniques, along with improved computational analyses, have significantly progressed the fundamental understanding of 3D nanomagnets and 3D chiral spin textures [31-33]. These developments have created fertile ground for exploring SW dynamics in these systems, driving the evolution of 3D magnonics.

Although the investigation of SW dynamics in 3D-ASI structures is still in its infancy, recent studies have shown promising results. One study on multilayered ASI nanoarrays, consisting of two magnetic layers separated by a non-magnetic spacer, demonstrates intriguing SW dynamics controlled by microstate, ultra-strong magnon-magnon coupling, and chirality-selective magnetic vortex states [34]. Moreover, micromagnetic simulations of magnetic buckyball nano-architectures reveal rich SW modes at vertices and nanowires, with properties that can be finely tuned through geometric alterations [35]. Another numerical study on the same system uncovers a complex thermal magnetic behaviour, characterized by five distinct crossovers that delineate six magnetic phases [36]. Starting from higher temperatures, the system transitions from a paramagnetic state to a spin-ice phase, eventually evolving into an imperfect charge-ordered crystal. Recent simulation results on 3D interconnected nanowire arrays also show SW modes generated by charged vertex states [37], analogous to effects observed in 2D-ASI [38]. The development, fabrication, and measurement of 3D-ASI, highlighting new frontiers in 3D model systems and applications is extensively discussed in a recent perspective [39]. Despite significant efforts to understand the dynamical properties of complex 3D systems, experimental results showing SW dynamics modulated by charged vertices in pure 3D-ASI are still lacking in the literature. Additionally, experimental investigations into magnetic configuration-driven SW dynamics in 3D-ASI are also scarce. This research gap can be addressed by expeimental study on 3D-ASI structures with nanowires arranged into a diamond-bond 3D lattice geometry. In recent years, this system has demonstrated a wide range of rich phenomena, including magnetic charge propagation [40] and the excitation of coherent SWs [41]. Additionally, a recent Monte-Carlo study of this structure has revealed an exotic phase diagram consisting of the conventional ice-phase and monopole crystals consisting of single- and double-charged tiling. Magnetic imaging of experimental structures after in-plane demagnetization found crystallites of single magnetic charge superimposed upon an ice background [42]. This intriguing 3D system offers an exceptional opportunity to explore SW dynamics, which can be tuned by magnetic charged states and configurational anisotropy.

In this article we directly demonstrate the detection of magnetic charged vertices in a 3D-ASI through measurement of SW dynamics using conventional Brillouin light scattering (BLS) and supporting micromagnetic simulations. Zero-field SW spectra measured for two sets of samples, with and without magnetic charged vertices, reveal an intrinsic difference in SW excitation characteristics. Furthermore, we explore the configurational anisotropy-dependent tunability of SWs, which highlights the significant variations in the number and field-evolution of SW modes depending on the direction of external applied field (*H*). These findings are key to understanding the dynamics of complex 3D systems and advancing reprogrammable 3D magnonic crystals for practical applications.

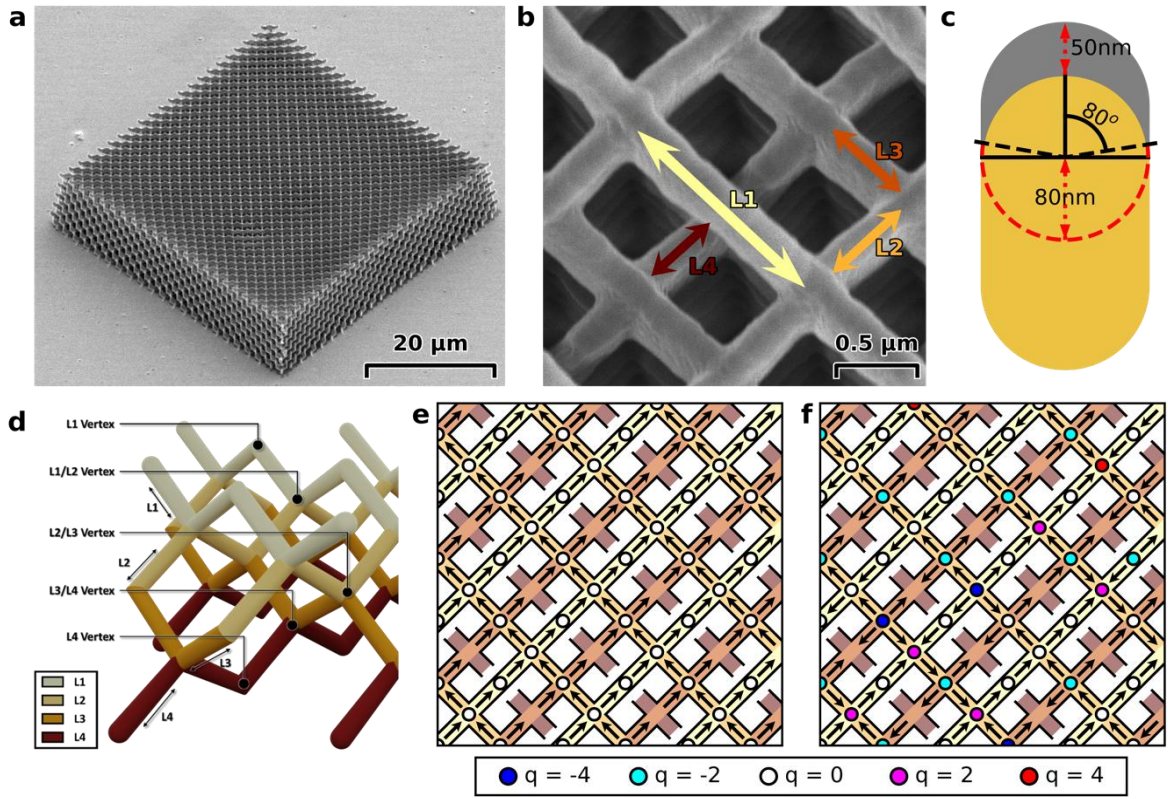

**Fig. 1| Static characterization of 3D-ASI. a** Scanning electron microscopy image of a representative 3D-ASI sample, which takes the form of a diamond-bond lattice **b** Zoomed image showing the individual sub-lattices. **c** Schematic of the nanowire cross-section. Magnetic material (grey) takes a crescent-shaped cross-section, upon the underlying polymer structure (yellow). **d** A unit cell of the 3D-ASI is shown schematically, with sub-lattice coloured differently according to legend **e** Arrow map showing the measured magnetization configuration of the magnetized sample, as based upon magnetic force microscopy (MFM) measurements (See Extended Data). Circles at the vertex indicate magnetic charge as described in the legend and individual sub-lattices are coloured according to legend in **d**. The sample has a two-in/two-out, T2 tiling at every vertex. **f** Arrow maps showing the measured magnetization configuration of the demagnetized sample, based upon MFM. The sample is found to possess charge-ordered regions, superimposed upon an ice T2 background.

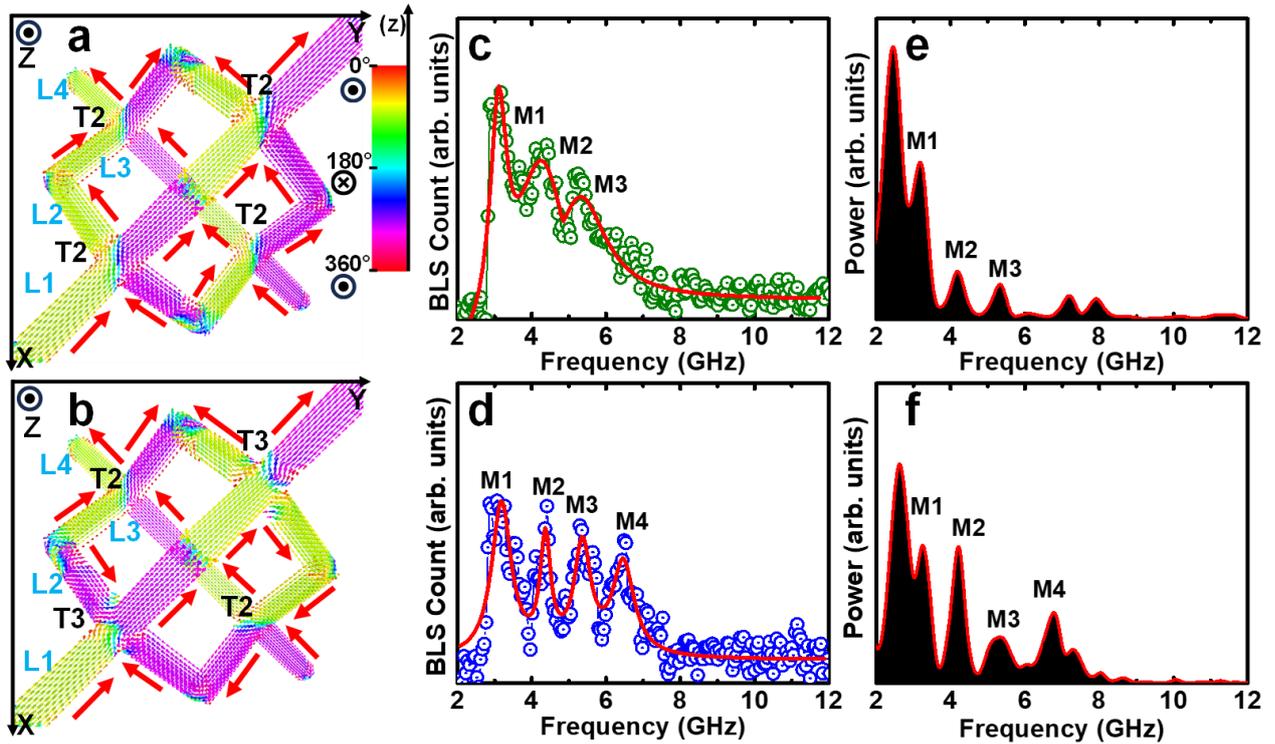

**Fig. 2| SW dynamics of magnetized and demagnetized samples.** Ground-state magnetic microstate representing the unit cell of diamond bond lattice for the magnetized **a** and demagnetized **b** sample, where the red arrows indicate the spin direction. The colour of the spins represents their orientation relative to the z-axis, as indicated by the colour bar located on the right side of **a**. The sublattices (L1, L2, L3 and L4) have been indicated in light blue colour and the vertex type (T2 and T3) in black colour. Experimental BLS spectra highlighting the presence of three SW modes in the magnetized sample **c** and four SW modes in the demagnetized sample **d**. The simulated SW spectra from the simulated unit cell of magnetized sample **e** and demagnetized sample **f**. Open circles in **c** & **d** represent the experimental data points, while the red and pink solid lines correspond to the results of the multipeak fitting.

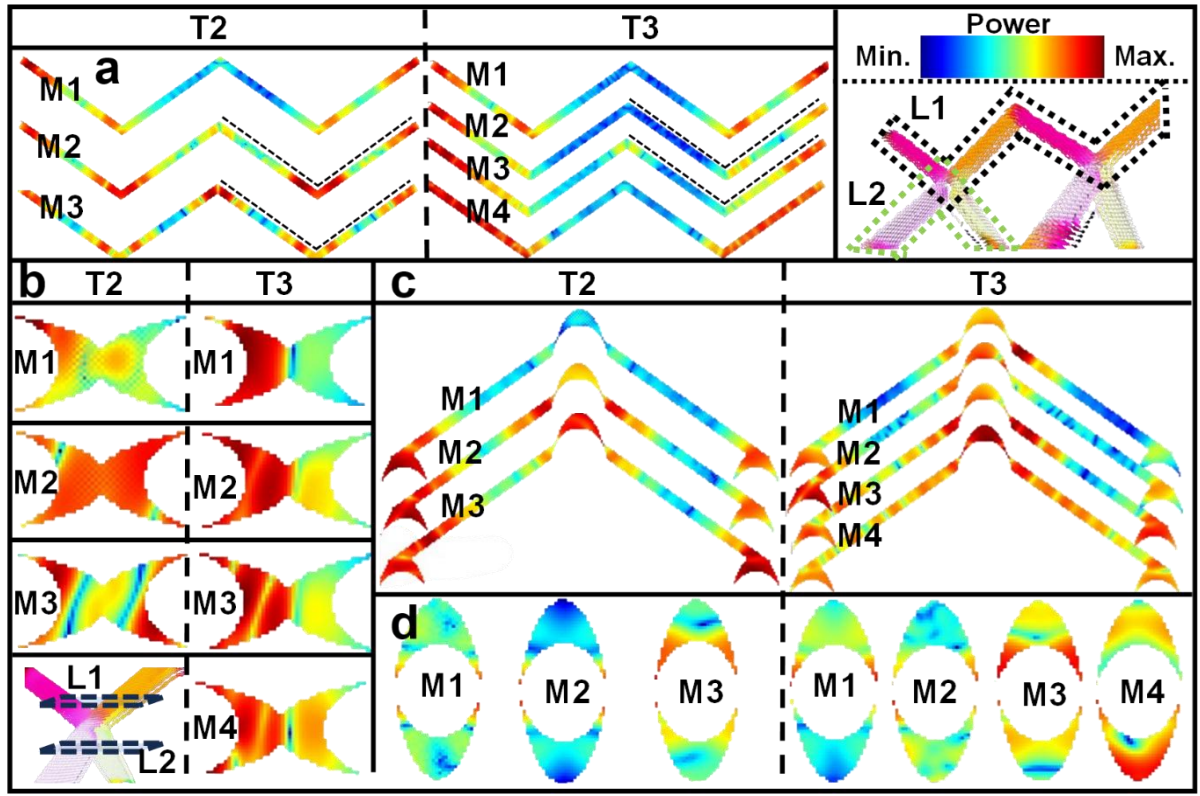

**Fig. 3| SW excitation characteristics of 3D-ASI nanowires and vertices.** SW power profiles of the L1 sublattice along the length **a** and cross-section **b** with T2 and T3 vertices, along with the schematic illustrations. SW power distributions of the L2 sublattice along its length **c** and cross-section **d**, featuring T2 and T3 vertices.

## Magnetic Charge Driven Spin-Wave Dynamics

Geometrically frustrated ASI systems exhibit intricate SW dynamics due to the energy differences between various macrospin configurations, with these dynamics strongly dependent on the magnetic microstates in 2D [3,43,44]. One of the intriguing features of ASI is its ability to tune the magnetic microstate through magnetization and demagnetization processes [42,44]. Leveraging this property, we fabricated 3D-ASI samples using a combination of two-photon lithography and evaporation (See Methods), which were subject to distinct magnetic field protocols. A scanning electron microscope (SEM) image of a typical lattice with array dimensions of 50×50×10 μm³, is shown in Figure 1a, with individual sub-lattices shown in Figure 1b. Individual nanowires (length = 866 ± 10 nm) have a crescent shaped cross-section as shown schematically in Figure 1c. A 3D perspective of a unit-cell is shown in Figure 1d, allowing the depth of each sub-lattice to be identified.

The magnetized sample was created by first applying and then reducing $H$ along the surface termination sublattice (L1, Figure 1b). The demagnetized sample was obtained by spinning the 3D-ASI array in an oscillating magnetic field $H$ (details in the methods section), similar to protocols carried out in 2D-ASI. Our previous work using MFM revealed distinct differences in vertex types: the magnetized sample [40,41] contains only type-II (T2, charge-free) vertices, while the demagnetized sample exhibits magnetic charge crystallites superimposed upon an ice background. Overall, this yielded primarily coordination-number four vertices of type-III (T3) and T2 in an approximate ratio of 2:1, with few type-I (T1) and type-IV (T4) vertices [42]. The samples studied here show qualitatively similar behaviour as can be seen in the magnetization profiles for magnetized and demagnetized samples shown in Figure 1e and 1f, respectively. Here one can see that the magnetized sample consists only of T2 vertices and the demagnetized sample has regions with crystallites of magnetic charge (T3), with alternating positive (pink) and negative (blue) charges on L2. Note that T4 vertices with a magnetic charge, $q = \pm 4$, are only rarely observed. The associated atomic force microscopy (AFM) and MFM can be found in the Extended Data Figure 1.

The SW dynamics of these samples was investigated using the BLS technique (see Methods section). The measurement geometry is illustrated in Supplementary Figure S1. Micromagnetic simulations using the GPU-accelerated micromagnetic simulation package mumax$^3$ (see Methods) provide insight into the experimental results. The micromagnetic simulations were conducted on a unit cell of the diamond bond lattice. The charge distribution within the simulated unit cell was inspired by the real charge configurations observed in previously studied magnetized and demagnetized samples [40-42], as shown in Figures 2a and 2b, respectively.

The modulation of SW excitation, driven by the distinct energy landscapes associated with charged and uncharged vertices [40,42], is clearly evident from the measured BLS spectra of the magnetized and demagnetized samples (Figures 2c and 2d, respectively). In the magnetized sample, three SW modes are observed, labelled M1, M2, and M3. In contrast, an additional SW mode, labelled M4, emerges in the demagnetized sample. This result provides a robust experimental evidence of magnetic charge-dependent SW excitation in a pure 3D nanomagnetic structure, a concept that was previously confined to simulation results alone [35,37]. Our simulated results closely align with the experimental findings, successfully reproducing the key outcomes for both the magnetized and demagnetized samples, as shown in Figures 2e and 2f, respectively. However, minor discrepancies, such as the prediction of additional modes and differences in mode resolution, are unavoidable. These variations arise from intrinsic differences between the simulations and experiments, including sample morphology, measurement methods, and excitation techniques, as previously noted for 3D systems in earlier studies [28,41].

The SW excitation characteristics in the magnetized and demagnetized samples differ not only in the number of SW modes but also in their inherent spatial excitation patterns as shown in the spatial power profiles in Figure 3. These were obtained by analyzing the dynamic magnetization components using the custom-built MATLAB-based code DOTMAG [45]. Figure 3a shows the spatial power distribution of the SW modes along the L1 sublattice. Notably, for the T2 vertex, the M2 and M3 modes exhibit a symmetric power distribution in the adjacent arms (indicated by the dotted line). In contrast, for the T3 vertex, a distinct asymmetry in power distribution emerges between the two arms for the same modes. The power asymmetry between the two arms of the L1 sublattice for the T3 vertex is also evident in the cross-sectional power distribution profile shown in Figure 3b whereas for the T2 vertex of the same sublattice, it is symmetric. The spatial power profile of the SW modes along the L2 sublattice, shown in Figure 3c, illustrates that most SW modes in T2 and T3 are localized, with their spatial power concentrated around the vertex. However, the power of the M4 mode in the T3 vertex is more distributed along the arms of the tetrapod. The cross-sectional power distributions for the L2 sublattice (Figure 3d) in both the T2 and T3 vertices are symmetric. It demonstrates a common trend that the spatial power is concentrated at the edges and inner surfaces for low-frequency SW modes, and as the frequency increases, the power enhances inside the cross-section. The phase profiles of SW modes in the L1 and L2 sublattices, showing the diverse quantization characteristics of the T2 and T3 vertices, are discussed in Supplementary Section II. A comparative analysis of the power and phase profiles in the L3 and L4 sublattices with T2 vertices, for both magnetized and demagnetized samples, is detailed in Supplementary Section III. The SW excitation characteristics of the T1 and T4 vertices at the L1 sublattice is discussed in Supplementary Section IV, highlighting the symmetric power distribution between adjacent arms.

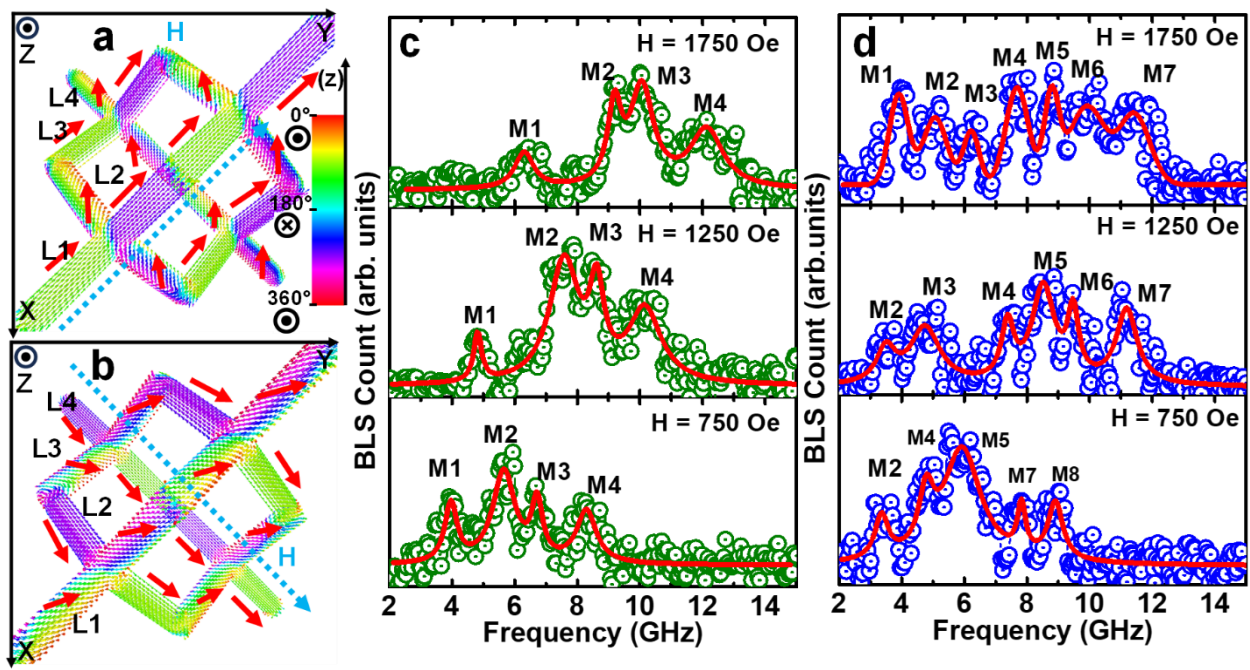

**Fig. 4| BLS spectra influenced by strength and direction of *H*.** Ground-state magnetic microstate at *H* = 750 Oe when *H* is applied parallel **a** and perpendicular **b** to L1, where the red arrows indicate the average spin direction. The colour of the spins represents their orientation relative to the z-axis, as indicated by the colour bar located on the right side of **a**. The sublattices have been indicated in black colour and the direction of *H* is in light blue colour. The experimental BLS spectra revealing the presence of SW modes at *H* = 1750 Oe, 1250 Oe, and 750 Oe, when *H* is applied parallel **c** and perpendicular **d** to the L1 sublattice. Open circles represent the experimental data points, while the red and pink solid lines correspond to the results of the multipeak fitting.

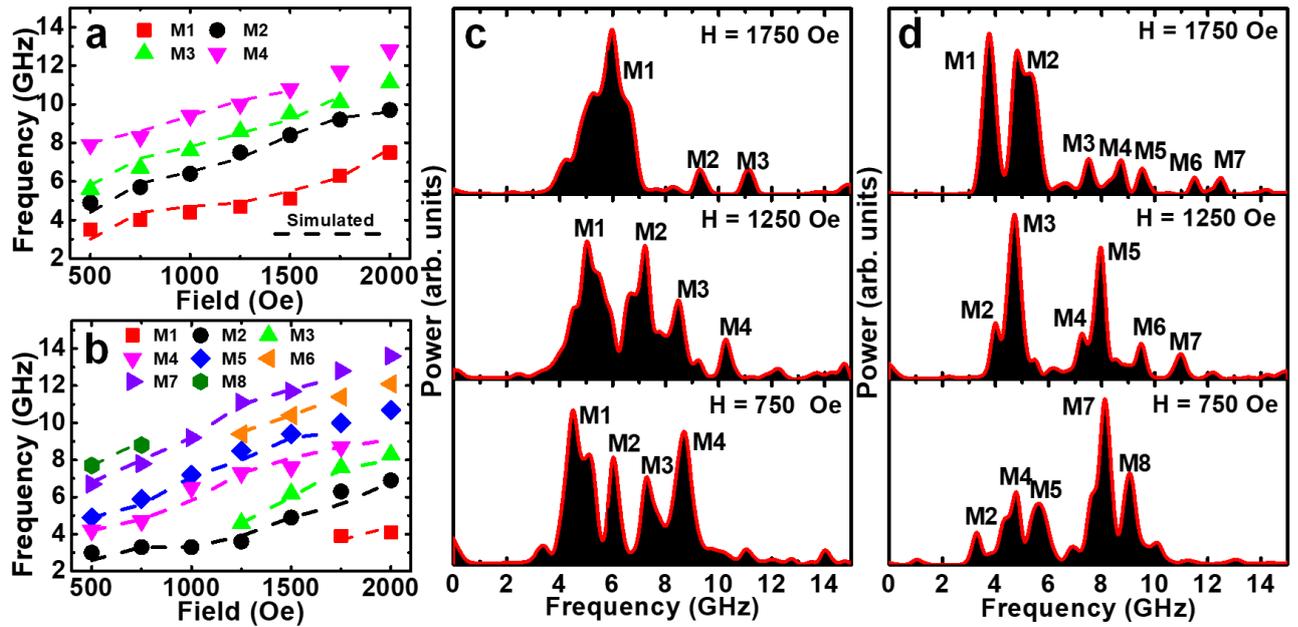

**Fig. 5| Configurational anisotropy dependent SW field dispersion.** SW field dispersion measured for two different $H$ directions, both parallel **a** and perpendicular **b** to L1. Simulated SW spectra obtained at $H$ = 1750 Oe, 1250 Oe, and 750 Oe when $H$ is applied parallel **c** and perpendicular **d** to L1.

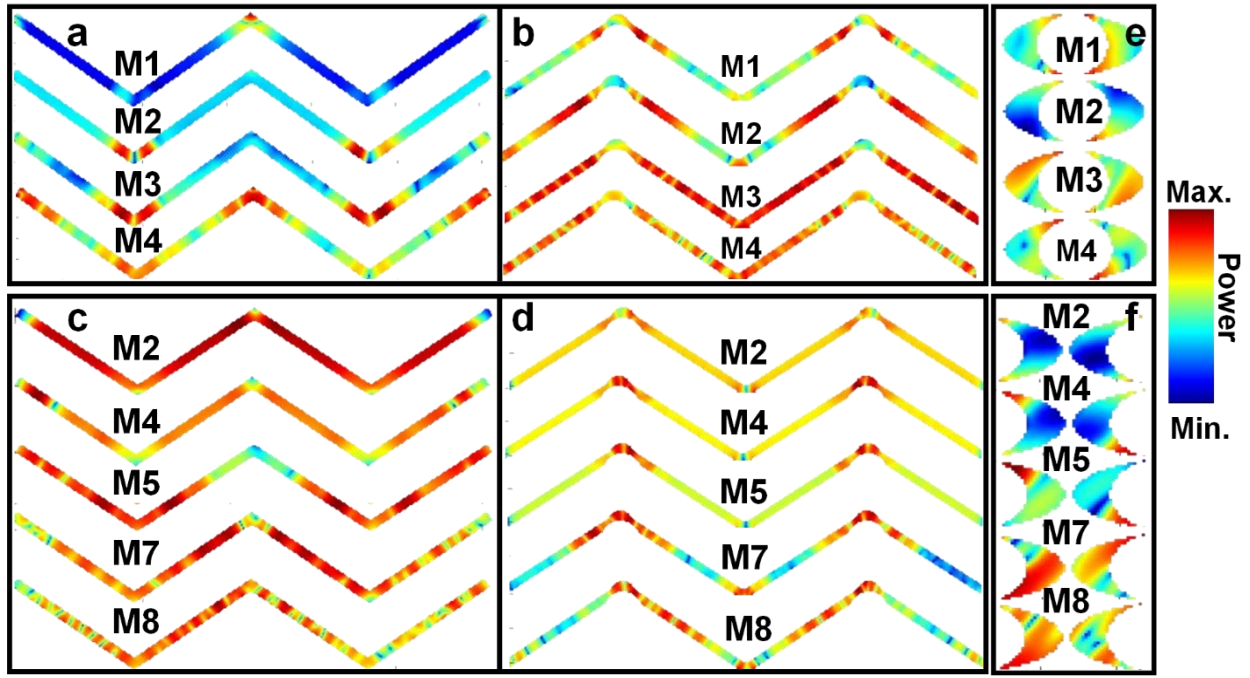

**Fig. 6| Spatial power profile of the SWs.** SW power profile (at $H$ = 750 Oe) along the L1 and L4 sublattices, when $H$ is along L1 (**a,b**) and perpendicular to L1 (**c, d**). The cross-section power profile of the SCS L4 **e** when $H$ is along L1 and SCS L1 **f** when $H$ is perpendicular to L1.

**Configuration Modulated Spin-Wave Dynamics**

Subtle changes in the direction of *H* lead to significant variations in configurational anisotropy, drastically altering the internal field distribution and resulting in intricate SW dynamics in magnonic crystals in 2D [46-49] including 2D-ASI [48,50-52] and in 3D [28,37]. To gain a deeper understanding of the configurational anisotropy-modulated SW dynamics in this complex 3D nanomagnetic structure, we have explored the field-dispersion along two different directions, i.e. by applying *H* parallel and perpendicular to the surface termination sublattice L1, as depicted in Figures 4a and 4b, respectively. The field-dispersion measurements were conducted from high field to low field, specifically from 2 kOe to 0.5 kOe, in intervals of 0.25 kOe. The BLS-spectra at three *H* values (750, 1250, and 1750 Oe) for the cases when the *H* is applied parallel and perpendicular to L1 are shown in Figures 4c and 4d, respectively. A clear distinction in the number of SW modes is visible in the BLS spectra obtained for the two different directions of *H*. Four modes are observed for all three different values of *H*, when the bias field is applied parallel to L1. However, for the same values of *H*, the number of modes significantly increases when the direction of the *H* is perpendicular to the L1 sublattice. The measured field-dispersion reveals a remarkable difference in SW characteristics on tuning the direction of *H*. When the *H* is applied parallel to the L1 sublattice, four consistent modes (shown in Figure 5a) are observed throughout the variation of *H* from 2.0 kOe to 0.5 kOe. In contrast, when the *H*-direction is switched perpendicular to the L1 sublattice, the field-dispersion exhibits additional intriguing features along with the presence of four consistent modes (shown in Figure 5b). These include the disappearance of the lowest frequency mode M1 below 1.75 kOe, merging of modes M2 with M3 and M5 with M6 at 1.25 kOe, and splitting of mode M7 to generate a high-frequency mode M8 around 1.0 kOe.

The variation of SW characteristics with the *H*-direction can be attributed to the change in the internal field profiles due to the alteration of the demagnetizing field, as explained for the single crescent-shaped structure in a recent publication [53]. However, to better understand the intriguing features observed in different fields, micromagnetic simulations were utilized (see Methods). For this purpose, the field dispersion is analyzed both parallel and perpendicular to the L1 sublattice of the unit cell (Figure 5a and 5b, dashed curve). All the key features observed in the experimental results are well reproduced by the micromagnetic simulations. However, minor discrepancies, such as the prediction of some extra modes and differences in the resolution of modes (as shown in Figure 5c and 5d), are unavoidable due to intrinsic differences between the simulation and the experiment for 3D systems, as described in detail in the literature [28,41].

To understand the *H*-dependent dynamic properties of the SWs, the simulated field evolution of the equilibrium magnetic microstate is analyzed (shown in Extended Data Figure 2). This analysis reveals significant differences in the microstates depending on the direction of *H*. When *H* is applied parallel to the L1 sublattice, a component of *H* aligns with the easy axis of the L1 and L3 sublattices, resulting in the alignment of the magnetic moments along the axis of the wire in the direction of *H*. In contrast, for the L2 and L4 sublattices, *H* aligns with the hard axis of the wire, which has a non-trivial crescent-shaped cross-section, leading to an unsaturated state due to the competition between shape anisotropy and *H*. This unsaturation results in the canted alignment of the spin in the wire, as shown in Extended Data Figure 2. The configuration is reversed when *H* is applied perpendicular to the L1 sublattice, as *H* is now along the hard axis for L1 and L3, and a component of *H* aligns with the easy axis of the L2 and L4 sublattices. The demagnetizing field distribution, as shown in Supplementary Figure S8, reveals that the demagnetizing field is predominantly present in the unsaturated sublattices. Consequently, these sublattices can be described as 'spin-canted sublattices' (SCS). In contrast, the demagnetizing field is comparatively weak in the sublattices along the $H$-direction and, therefore can be referred to as 'spin-aligned sublattices' (SAS).

The SCS play a dominant role in the excitation of the SW modes, as evidenced by the power profile of the SW modes. Figures 6a and 6b show the power concentration of the SW modes in the SAS L1 and SCS L4, respectively, at $H = 0.75$ kOe (along L1). These figures highlight that the dominant excitation of SWs occurs in the SCS, especially for the low-frequency modes. As the mode frequency rises, the power gradually increases in the SAS. Nevertheless, the primary excitation remains in the SCS. The primary excitation of the SWs in the SCS aligns with the change in the direction of $H$, as shown in Figures 6c and 6d. These figures demonstrate that SW modes have greater power in the L1 sublattice compared to the L4 sublattice. The cross-sectional power concentrations of the sublattice L4 ($H$ parallel to L1) and L1 ($H$ perpendicular to L1) are shown in Figures 6e and 6f, respectively. These show that the low-frequency modes are mostly excited at the edges, and with increasing frequency, the excitation gradually shifts towards the interior of the cross-section. These features of SW excitation align with the variation of $H$, as detailed in Supplementary Section V. Additionally, the phase profile of SW modes at $H = 750$ Oe, discussed in Supplementary Section VI, demonstrates the dominant quantization of SW modes in the SCS.

The field evolution of the magnetic microstates, as shown in Extended Data Figure 2, illustrates that as the field decreases, shape anisotropy becomes dominant and attempts to align the spin along the direction of the wire. The SAS remain magnetized throughout the field variation. However, the SCS exhibit peculiar characteristics with a decreasing field. The field evolution of the demagnetizing fields in the SCS L4 and L1, when $H$ is applied parallel and perpendicular to L1, respectively, is discussed in Supplementary Section VII. This illustrates that the field evolution of the demagnetizing field is highly monotonic when $H$ is applied parallel to the L1 sublattice, whereas it is non-monotonic when $H$ is perpendicular to L1. This suggests that the variation in SW field-dispersion with the direction of $H$ can be attributed to the differing field evolution of the SCS.

**Conclusion**

In conclusion, we demonstrate direct fingerprints of magnetic charged states in the SW dynamics by exploring 3D-ASI systems with and without magnetic charged vertices using conventional BLS spectroscopy and micromagnetic simulations. The BLS spectra obtained from the demagnetized sample, which contains magnetic charged vertices, show the presence of an additional SW mode as opposed to the BLS spectra from the charge-free magnetized sample. This additional SW mode is clearly due to the presence of the charged vertices, a finding also supported by micromagnetic simulations and consistent with magnetic force microscopy of the sample. Furthermore, the simulated SW power profile reveals that this mode exhibits intrinsic variations in SW excitation, demonstrating an extended nature, whereas most of the other SWs are localized.

Field-dependent investigations of SW dynamics reveal significant variations in the experimental field-dispersion of SW modes, including variations in the number of modes and their field evolution characteristics, which are influenced by the direction of the applied field w.r.t. the lattice. Alongside supporting the experimental findings, micromagnetic simulations reveal the presence of 'spin-aligned sublattices' (SAS) and 'spin-canted sublattices' (SCS) depending on the field direction. The SCS, characterized by canted magnetic moments due to the field applied along the hard axis, significantly influence SW excitation, as emphasized by the simulated SW power profile. Our findings suggest that direction-dependent field-dispersion can be attributed to the varying distribution and field evolution of demagnetizing fields associated with the SCS. Our unique results, showcasing the dynamic tunability of SW through magnetic charges and configurational anisotropy, serve the dual objective of enhancing the fundamental understanding of these intricate 3D nanomagnetic networks and exploring new possibilities for the development of reprogrammable 3D magnonic and spintronic devices.

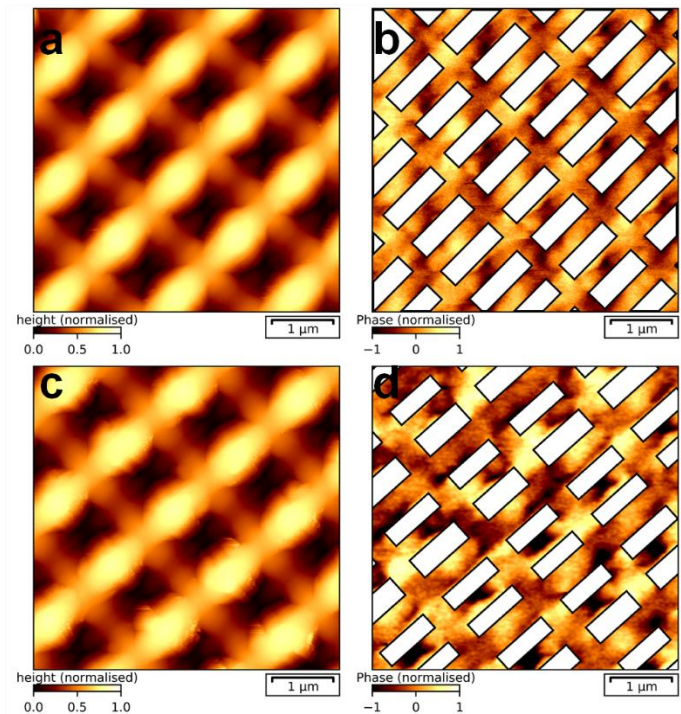

**Extended Data Fig. 1| Magnetitic Force Microscopy.** AFM and MFM image taken at remanence state for the magnetized **(a,b)** and demagnetized sample **(c,d)**.

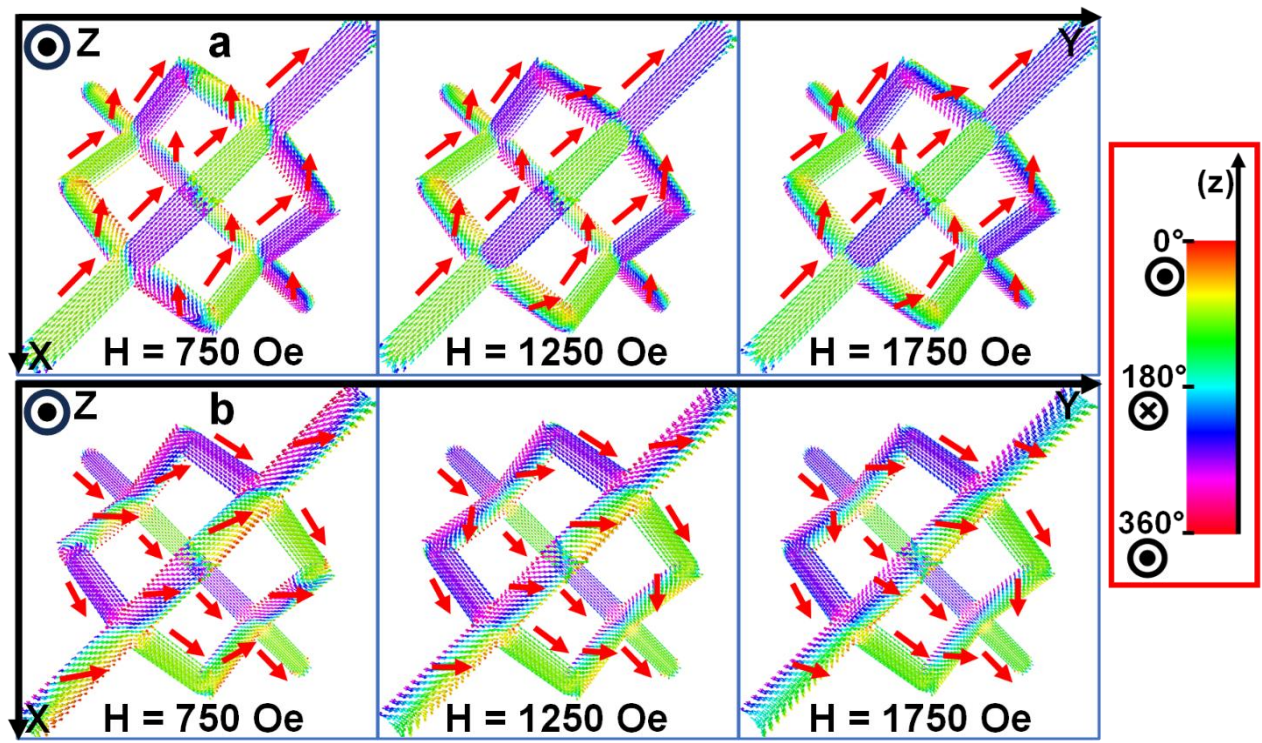

**Extended Data Fig. 2| Field evolution of magnetic microstate.** The simulated ground-state magnetic microstate at $H$ = 750 Oe, $H$ = 1250 Oe and $H$ = 1750 Oe, when $H$ is parallel **a** and perpendicular **b** to L1. The red arrows indicate the average spin direction. The colour of the spins represents their orientation relative to the z-axis, as indicated by the colour bar located on the right side of figure.

## Methods

### Sample Fabrication and Characterization

The 3D-ASI lattices were produced using TPL involving a 780-nm femtosecond pulsed laser focused to a diffraction limited spot where a photoresist is polymerized. Using a set of galvo mirrors, the focal spot is translated through the resist to trace the desired geometry. The glass substrate was prepared with a 20-minute acetone bath in an ultrasonic cleaner followed by a 20-minute isopropanol bath in the ultrasonic cleaner and subsequently dried using compressed air. Immersion oil was drop-cast on one side of the coverslip along with a negative-tone photoresist (IPL, proprietary to NanoScribe GMBH) drop-cast on the reverse side. The coverslip is loaded into the TPL system, and a script outlining the desired geometries is executed. After exposure, the samples were placed in a propyl glycol monomethyl ether acetate (PGMEA) developer bath to remove unexposed resist, followed by an isopropyl alcohol (IPA) bath and then gently dried using compressed air.

Four 15-nm gold layers were deposited using thermal evaporation at a 20° angle, with the sample stage rotated 90° between each evaporation. Finally, 0.067g of permalloy ($Ni_{81}Fe_{19}$) was deposited to yield a 44-nm permalloy layer. The base pressure for all evaporations was below $10^{-6}$ mbar. SEM was carried out using a Hitachi Regulus 8230 SEM.

MFM measurements were performed using the Bruker Dimension ICON scanning probe microscope in tapping mode using supersharp ultra-low moment probes from NANOSENSORS magnetized using a 0.5 T permanent magnet. The samples were positioned such that the L1 sublattice is parallel to the cantilever, and the scan direction set to 45°. The lift mode scans for obtaining magnetic contrast were performed at a 130-nm lift hight, allowing for higher drive amplitudes for improved magnetic contrast.

<u>Magnetization and Demagnetization</u>: The magnetized sample was obtained by applying an in-plane magnetic field, of magnitude 100 mT along the projection of L1. For demagnetized sample, we used a protocol typically used in 2D-ASI, based on method 1 outlined in prior work [54]. An in-plane oscillating field is applied to the sample, starting at 0 mT and ramping up to 75 mT at 2.5 T s$^{-1}$, the amplitude of the oscillation decays to 0 mT over five days. Effective field rotation is achieved by rotating the sample at 1000 rpm about the axis perpendicular to the field.

### Brillouin Light Scattering Spectroscopy

The SW dynamics of the 3D-ASI samples were measured using conventional BLS, a non-contact and non-invasive technique ideal for detecting thermally excited SWs at room temperature without external excitation. BLS operates on the principle of inelastic light scattering, described quantum mechanically as a photon-quasiparticle (magnon here) interaction, where the creation (Stokes process) and annihilation (anti-Stokes process) of magnons are observed. The BLS spectra were obtained in the Damon–Eshbach (DE) geometry using a Sandercock-type (3+3)-pass tandem Fabry-Pérot interferometer and a p-polarized single-longitudinal-mode solid-state laser (532 nm wavelength, 230 mW power). The laser power on the sample surface was 65 mW, with a spot size of ~50 μm in diameter, closely matching the sample's lateral dimensions [41]. This setup enabled SW measurements from nearly the entire sample volume. Cross-polarization between the inelastically backscattered and incident beams was used to minimize phonon interference. The sample was mounted on a 360° in-plane rotating stage, allowing precise rotation along the desired direction within the plane of the substrate. A permanent magnet was used to apply the $H$, and its strength was accurately measured at each step of the measurement to obtain the field-dispersion.

**Micromagnetic simulation**

The micromagnetic simulations were performed using the GPU-accelerated micromagnetic simulation package mumax[3] [55]. We utilized cuboidal cells with dimensions of 5×5×5 nm³, which is less than the exchange length of permalloy (≈5.2 nm). The material parameters used in the simulation were: gyromagnetic ratio ($\gamma$) = 17.6 MHz/Oe, saturation magnetization ($M_s$) = 860 emu/cc, anisotropy field ($H_K$) = 0, and exchange stiffness constant ($A_{ex}$) = 13×10$^{-7}$ erg/cm for permalloy [46]. To simulate the experimental charge configuration, we designed a unit cell of the diamond bond lattice composed of crescent-shaped nanowires with dimensions similar to the experimental sample and applied a 2D periodic boundary condition in the x−y plane. The simulated structure along the z-direction contains four sublattices, similar to the experimental sample. The magnetization of each wire was defined separately to mimic the charge configurations obtained experimentally for the magnetized and demagnetized samples and were relaxed to obtain the equilibrium magnetic microstate. For the simulation of SW dynamics, we applied a square-shaped pulsed magnetic field with a peak amplitude of 5 Oe along the z-direction, a rise and fall time of 10 ps each, and a duration of 20 ps to the equilibrium magnetic state using a Gilbert damping parameter ($\alpha$) = 0.008. The SW spectra were calculated by performing a fast Fourier transformation (FFT) of the x-component of the dynamic magnetization ($m_x$). To mimic the field-dispersion results, the external field was applied parallel and perpendicular to the L1 sublattice and allowed to relax to obtain the equilibrium magnetic microstates. Thereafter, the SWs were excited using the same technique mentioned above, except that the amplitude of the square-shaped pulsed magnetic field was increased to 20 Oe.

**Data availability**

The datasets obtained and/or analyzed in this study are available from the corresponding author upon reasonable request.

**Code availability**

The code used in this study are available from the corresponding author upon reasonable request.

## Acknowledgements

AB acknowledges the financial assistance from the Nano Mission, Department of Science and Technology, Government of India under Grant Nos. DST/NM/TUE/QM-3/2019-1C-SNB and SR/NM/NS-09/2011. CK and AKM acknowledge S. N. Bose National Centre for Basic Sciences, SM acknowledges University Grants Commission and SP acknowledges Council of Scientific & Industrial Research for their respective fellowship. SL acknowledges funding from the Engineering and Physical Sciences Research Council (EPSRC), grant number EP/X012735/1 and the Leverhulme Trust, grant number RPG-2021-139.

## Author contributions

AB and SL planned and supervised the project. AvDB prepared the samples and made the AFM and MFM measurements. CK, AKM, SP and SM made the BLS measurements. CK made the micromagnetic simulations and data analyses in consultation with AB, AvDB and SL. JRS and AOA made some test BLS measurements and participated in the discussion. CK, AB and SL wrote the manuscript in consultation with all authors.

## Competing interests

The authors declare no competing interests.

## Additional information

**Supplementary information** The online version contains supplementary material available at https://doi......................

**Correspondence** and requests for materials should be addressed to Anjan Barman or Sam Ladak.